# Protective Effect of Trehalose Sugar on Amyloid-Membrane Interactions using BLM Electrophysiology

## (Short title: Trehalose Protects Membrane from Amyloid)

Yue Xu[1], Carina Teresa Filice[2,3], Zoya Leonenko[1-3, *]

[1]Department of Physics & Astronomy, University of Waterloo, Waterloo, ON, N2L 3G1, Canada;

[2]Department of Biology, University of Waterloo, Waterloo, ON, N2L 3G1, Canada;

[3]Waterloo Institute for Nanotechnology, Waterloo, ON, N2L 3G1, Canada

*Email: zleonenk@uwaterloo.ca

## ABSTRACT

Alzheimer's disease (AD) is a neurodegenerative disease characterized by dementia and memory loss in the elderly population. The amyloid-β peptide (Aβ) is one of the main pathogenic factors in AD and is known to cause damage to neuronal cellular membranes. There is no cure currently available for AD, and new approaches, including preventive strategies, are highly desirable. In this work, we explore the possibility of protecting neuronal membranes from amyloid-induced damage with naturally existing sugar trehalose. Trehalose has been shown to protect plant cellular membranes in extreme conditions and modify Aβ misfolding. We hypothesize that trehalose can protect the neuronal membrane from amyloid toxicity. In this work, we studied the protective effect of trehalose against Aβ1-42-induced damage in model lipid membranes (DPPC/POPC/Cholesterol) using atomic force microscopy (AFM) and black lipid membrane (BLM) electrophysiology. Our results demonstrated that Aβ1-42 damaged membranes and led to ionic current leakage across these membranes due to the formation of various defects and pores. The presence of trehalose reduced the ion current across membranes caused by Aβ1-42 peptide damage, thus efficiently protecting the membranes. These findings suggest the trehalose sugar can potentially be useful in protecting neuronal membranes against amyloid toxicity in AD.

**STATEMENT OF SIGNIFICANCE**

Alzheimer's Disease (AD) presents a great challenge to scientists and society. There is currently no cure available for AD, and many approaches targeting Aβ have failed in clinical trials (1-4). As the disease progresses over a prolonged period, there arises a possibility to prevent or reduce its harm with preventive approaches. It has been shown that cellular membranes are primary sites of Aβ accumulation, and thus can be the target for prevention strategies. Our study explores a naturally extracted sugar, trehalose, as a potential agent for AD prevention and focuses on exploring the protective effects of trehalose on lipid membranes against Aβ-induced damage. Our findings may contribute to pharmaceutical research and the development of membrane-targeted preventive therapeutics for neurodegenerative diseases.

**INTRODUCTION**

*Amyloid β and Alzheimer's Disease.* Amyloid-β plaques are considered one of the main pathological features of Alzheimer's Disease (AD) (5-7). These plaques are attributed to the overproduction and aggregation of amyloid β (Aβ) peptides (8) and are found on the surface of neuronal cells (9-11). As the neuronal cell membrane is of critical importance in neural physiology (11) and is recognized as a target for amyloid attack (12-14), it provides an opportunity to explore new strategies to protect the neuronal cell from amyloid-induced damage (15).

It has been shown that small Aβ oligomers rather than large fibrils are most toxic to neuronal cells (16-19), with Aβ1-42 as the most toxic species (20, 21). The proposed mechanism of Aβ1-42 oligomer-mediated neurotoxicity suggests that Aβ1-42 binds to neuronal cellular membranes and causes membrane disruption by creating various defects, including ion channel-like pores, which lead to membrane damage, ion leakage and ionic imbalance (18, 22-24), thus contributing to the dysfunction, degeneration, and death of neurons (22, 25).

Due to the importance of cellular membranes (26, 27), they have attracted a lot of attention in connection to AD pathology. Previous studies suggest a strong correlation between the progression of AD and the alteration of lipid compositions in plasma membranes (15, 28-33). In addition, the reduction of sphingomyelin levels and elevation of cholesterol due to abnormal lipid metabolism was found in AD brains, presumably as a result of disease progression (33, 34). Such alteration in lipid composition significantly influences the interactions of membranes with amyloid (35-40). It has been shown by Drolle *et al.* that model lipid membranes which mimic healthy neuronal membranes are less susceptible to amyloid damage than membranes mimicking AD neurons (41). This evidence indicates a pertinent relationship between Aβ induced neurotoxicity and lipid membrane properties and suggests amyloids can recognize the changes in lipid membranes. As there is currently no cure available for AD and multiple drug candidates targeting amyloid aggregation are failing in clinical trials (1-4), it becomes important to explore potential preventive approaches, specifically targeting lipid membrane properties, as this may reduce amyloid neurotoxicity.

***Protecting Membranes Against Amyloid β Using Small Molecules.*** Consequently, inspired by this novel idea to focus on membrane protection strategies, current AD research has highlighted various small membrane-active molecules capable of blocking or reducing the insertion of Aβ oligomers in neuronal membranes, such as melatonin (42-44), curcumin (45), various aminosterols (46) etc. Specifically, we have demonstrated *in vitro* that melatonin reduces the binding of Aβ to model membranes, thus protecting the membranes from amyloid-induced damage (47). In studies completed by Limbocker *et al.*, aminosterols such as trudosquemine and squalamine bind to cellular membranes and displace toxic Aβ1-42 oligomeric species from neuronal cells in addition to modifying Aβ1-42 aggregation mechanisms in solution (48-50). Errico *et al.* expounded upon these findings by discovering that this aminosterol-mediated oligomer displacement from cell membranes occurs through the alteration of membrane properties such as lipid organization, lipid packing, and surface charge (46, 51). Other research groups have also reported neuroprotective effects of curcumin (45), tauroursodeoxycholate (52, 53), and 2-hydroxydocosahexaenoic acid (54) against Aβ toxicity by reducing binding and increasing membrane stability. Inspired by the membrane protection strategies, our studies aimed at exploring the influence of trehalose on Aβ-membrane interactions in AD.

***Trehalose, Aβ and Membranes.*** Trehalose is a non-reducing disaccharide of glucose often found in biological laboratories as an essential component of solutions used to preserve tissues, organs, and other biological samples (55). Trehalose can protect biomolecules (i.e. lipids, proteins) from extreme conditions such as high temperature and mechanical stress (55), with effective cryoprotection concentrations ranging from 2.6 μM (56) to 1M (57) depending on sample composition.

Additionally, trehalose, as a dietary sugar, shows little toxicity to humans, even with an average daily intake of 42.37 g from both natural sources and processed food (58). Most individuals can metabolize trehalose using the intrinsic enzyme trehalase; however, a small subset of the population may lack this enzyme, leading to side effects like diarrhea, discomfort, borborygmus, etc. which can be alleviated by oral ingestion of trehalase (59). In clinical ophthalmology, trehalose is an important component in novel topical dry-eye treatments that have shown improvements in animal models and clinical patients suffering from dry eye (60-62). Currently, trehalose is regarded as a bioactive agent and is also being explored for its therapeutical potency in diabetes, osteoporosis, and neurodegenerative diseases (58, 63). Animal studies have revealed the neuroprotection role of trehalose in neurodegenerative diseases correlating to the accumulation of aberrant proteins, such as Alzheimer's disease (AD), Parkinson's disease (PD), and Huntington's disease (HD) (63).

Previous studies suggest that trehalose can directly interact with membranes by replacing water molecules adjacent to membrane phospholipids and forming hydrogen bonds with the lipid headgroups (64-67). As cellular membranes play an important role in amyloid toxicity in AD, this provides an important avenue to explore the possibility of protecting the membrane and to prevent or reduce amyloid toxicity by modifying the membrane properties itself.

Trehalose is being investigated as a potential therapy for AD (68). In principle, we may consider two pathways where trehalose can protect against Aβ1-42 oligomers toxicity. The first pathway focuses on trehalose altering amyloid aggregation in solution. Previous studies by other research groups indicate that, over large periods of time, trehalose triggers the conformational change of Aβ and inhibits the formation of Aβ deposits in vivo (68-71). In a study conducted by Liu *et al*., trehalose showed different suppression effects on the aggregation of Aβ1-40 and Aβ1-42 respectively (69). The authors reported that the cytotoxicity of Aβ1-40 was reduced by inhibiting the formation of amyloid oligomers (69). Trehalose was also found to alter the conformation of Aβ1-40 oligomers from toxic β-sheets into the less toxic α-helix structure, thereby reducing amyloid insertion to membranes (67, 70). Conversely, the inhibition effect of trehalose on Aβ1-42 aggregation in solution was limited; while aggregation was reduced and fibril formation inhibited, the conformation of Aβ1-42 oligomers was not altered and therefore the toxicity was not fully eliminated (69). This proposed pathway of trehalose protection can be suppressed by the presence of salt (NaCl) at concentrations larger than 25 mM as indicated in studies observing trehalose interactions with similar β-sheet peptides lysozyme and insulin (72). Therefore, while trehalose shows a suppressive effect on fibril formation in solution, the toxic Aβ1-42 oligomers persist and thus continue to cause damage to membranes.

The second, less-studied pathway of trehalose protection focuses on protecting the membrane from amyloid damage. In addition to the reported effects of trehalose on plant (73) bio-membranes and amyloid proteins respectively (69, 71, 72), trehalose also reduced liposome permeability caused by Aβ1-40 oligomers (74). However, to our knowledge, most studies regarding incorporative behaviours of amyloid proteins and trehalose on lipid membranes concentrate on these aspects at the theoretical level alone (i.e., molecular dynamics simulations). Previous experimental studies investigated only Aβ1-40 rather than its more toxic counterpart, Aβ1-42. Therefore, the protective effect of trehalose against Aβ1-42 at the membrane level is not known and is the subject of this study.

We hypothesize that similarly to protecting the plant membranes against harsh conditions, the trehalose can bind to the surface of neuronal cellular membranes and protect them from toxic amyloid-induced damage (Figure 1). In this work, we aim to study the effect of trehalose on Aβ1-42 toxicity against model neuronal membranes composed of DPPC, POPC and cholesterol in a weight ratio of 4:4:2. Zwitterionic lipids with PC groups were chosen as the main constituent of our proposed simple neuronal lipid model as neutral-charged lipids constitute the main component of the cellular lipid membrane in neurons (41, 75), despite anionic lipid membranes being the focus of many Aβ-membrane studies (43, 67, 76). As shown in Figure 1, Aβ1-42 oligomers can bind to the cellular membranes and produce various defects (Figure 1(a)), increasing membrane damage. We propose that trehalose can protect the membranes against Aβ1-42 toxicity by binding to the membrane surface and by creating a protection barrier to resist amyloid-induced damage (Figure 1(b)).

The research focuses on studying membrane integrity during treatment with trehalose and damage from Aβ1-42 oligomers. We used the black lipid membrane (BLM) electrophysiology technique to investigate the protective effect of trehalose in model lipid membrane against damaging effects of Aβ1-42 by measuring membrane permeability, complemented by atomic force microscopy (AFM) used to visualize the amyloid aggregates and the damage induced by Aβ1-42 oligomers in the membranes. Our results demonstrate that high concentrations of trehalose improve membrane stability after the destruction of Aβ1-42 to the membrane and reduce the damage induced by toxic Aβ1-42 oligomers.

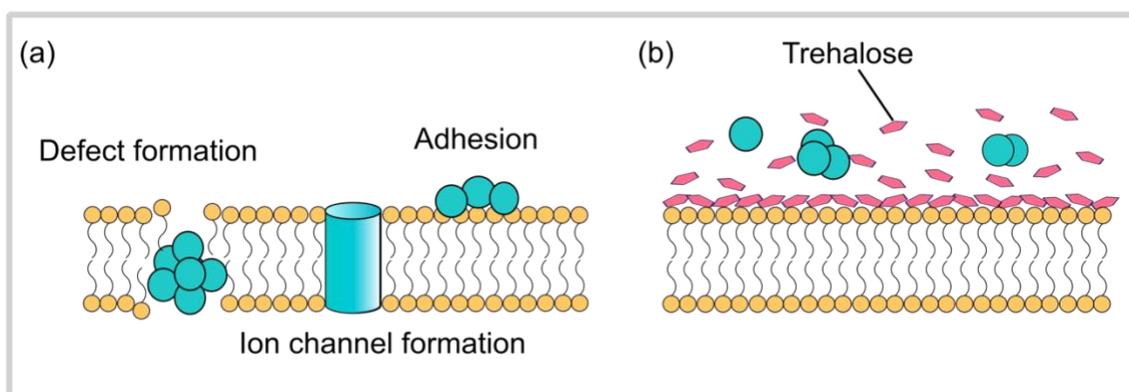

**Figure 1. Hypothesis of trehalose protection mechanism against amyloid damage to lipid membranes.** (a) Multiple mechanisms of Aβ1-42 damaging lipid membranes. (b) A potential mechanism that trehalose protects membranes against Aβ1-42 toxicity.

**METHODS**

**Black lipid membrane (BLM) electrophysiology.**

BLM is a popular electrophysiology method used to measure the permeability of model lipid membranes, where membranes are suspended over a micrometer-sized pore, and ion current is measured as an indication of the membrane permeability, directly related to membrane damage (Figure 2).

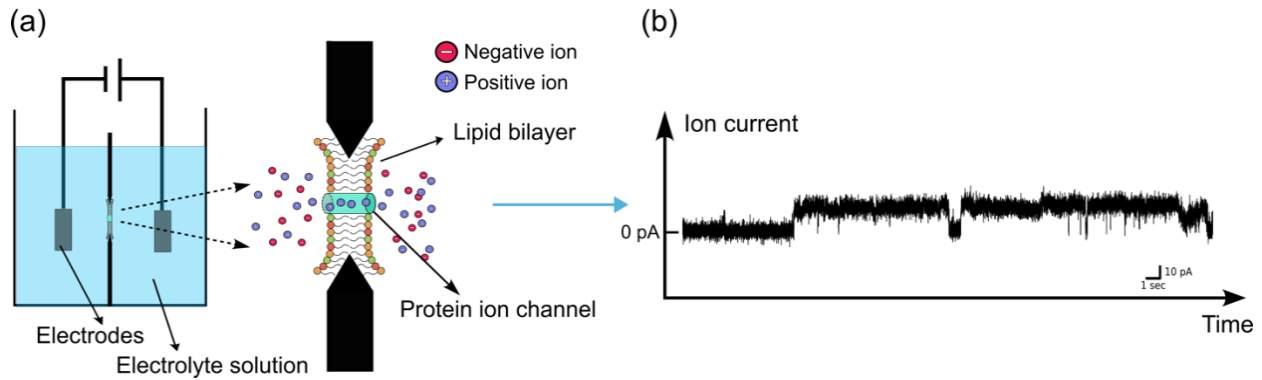

**Figure 2. Schematic of electrophysiological planar bilayer recording system** used in (a) BLM technique to detect protein ion channels in lipid bilayers and (b) a representative current trace of Aβ1-42 ion channel forming in suspended lipid bilayers in KCl-HEPES buffer at 120 mV, obtained from single-event recording experiments.

***Lipid and model membrane preparation for BLM.*** Model membranes were composed of 1,2-dipalmitoyl-sn-glycero-3-phosphocholine (DPPC), 1-palmitoyl-2-oleoyl-sn-glycero-3-phosphocholine (POPC) and cholesterol (Sigma-Aldrich, Canada) in mass ratio 4:4:2. Powdered lipids were dissolved in 9:1 chloroform and methanol (v/v) to prepare a stock solution at a concentration of 10 mg/mL, which was stored at -20 °C. To create a working lipid solution for BLM experiments, an appropriate amount of 10 mg/mL lipid stock in chloroform and methanol (9:1v/v) was evaporated by $N_2$ gas until completely dried into a film. Then, n-decane and butyl alcohol (50:1 v/v) were added to resuspend the lipid solution to a final concentration of 10 mg/mL.

***Amyloid preparation for BLM.*** Amyloid solution was added to the liquid chamber with suspended membrane at concentration of 110 μM. Aβ1-42 (Ultra-Pure, NaOH, rPeptide, USA) was dissolved at 0.5mg/mL in 150 mM KCl and 20 mM HEPES buffer (final concentration 110 μM). Aβ1-42 solution was distributed into aliquots and stored at -20°C. At these conditions, we obtained small soluble oligomers. Prior to addition in BLM, these pre-dissolved aliquots were sonicated for 30 second to 1 minute.

***BLM recordings***. We used BLM instrument from Elements (Italy). Prior to lipid membrane formation, a 1.3mL Delrin cuvette and chamber were filled with electrolyte solutions. Potassium chloride (KCl) was used as the electrolytes, given its supreme conductivity. The electrolyte solutions in our experiments were (a) 150mM KCl and 20 mM HEPES buffer, (b) with 50 mM trehalose, and (c) with 100 mM trehalose respectively. Ag/AgCl electrodes were placed in reservoirs on either side of the cuvette aperture to record the current change across the suspended membrane over time. To create a suspended lipid membrane, the working lipid solution was painted over the 150 μm aperture in the Delrin cuvette using a shaved pipette tip. Once a stable membrane was formed, Aβ1-42 solution was added into the chamber to a final concentration of 5 μM. Measurements were conducted with BLM instrument with eONE amplifier (Elements, Italy), and real-time recording was monitored by Element Data Reader (EDR) software. The sampling rate (SR) was set as 5 kHz and a filter was applied to the real-time trace to generate a final bandwidth of SR/20. The applied current across the membrane was recorded at a range of constant voltages selected based on findings from voltage dependence tests described below.

*1. Voltage Dependence Test.* The voltage dependence method measures the integral membrane conductance before and after the formation of stable ion channels, which were able to stay in the membrane at least 2 min and evaluate the destruction level of membranes (77, 78). Currents were recorded in a series of stepwise voltages from -120 mV to + 120 mV in increments of 10 mV. Then, membrane conductance was obtained by I-V plot in terms of Ohm's Law $C = \frac{I}{V}$, where I is recorded current, and V is applied voltage (79). At least 6 effective I/V curves were collected for each experiment condition.

*2. Single-event ion current recording.* Single-event ion currents were recorded over time to study the acute and instantaneous membrane current fluctuation due to peptide-membrane interactions. Large current changes induced by protein-membrane interaction were observed and recorded in the voltage range of 60 to 175 mV. This method evaluates the magnitude of single ion current events caused by interaction of Aβ1-42 with membranes (79). Single events for each treatment were collected from at least 10 effective traces with observable current increase due to amyloid insertion.

**Data acquisition and statistics for BLM.** Data that included currents and corresponding voltages were obtained by EDR software and were processed using Clampfit 11.2 software. Figure generations, data analysis, and hypothesis testing (one-way ANOVA, Tukey's test) were performed in OriginPro 2020.

**Atomic force microscopy (AFM)**

Atomic force microscopy is a high-resolution imaging technique capable of generating nanometer-scale resolution images of biological samples in physiologically relevant conditions (80). We used tapping mode (intermittent contact mode) AFM (81) imaging in air to evaluate the size of Aβ1-42 oligomers and in liquid to visualize the membrane and Aβ-membrane interactions.

*Aβ1-42 oligomers preparation for AFM.* The method of Aβ solution preparation for AFM is similar to that for BLM measurement. Aβ1-42 (Ultra-Pure, NaOH, rPeptide, USA) was dissolved in 150 mM NaCl and 20mM HEPES buffer to a final concentration of 110 μM. The Aβ-buffer solution was distributed into aliquots and stored in -20˚C. Prior to addition in liquid cell, aliquot was sonicated for 1 minute to reach room temperature and to ensure the protein uniformly distributed in solution.

Aβ1-42 oligomers dry sample is prepared for AFM imaging in air. 10 μL Aβ-buffer solution was deposited onto the freshly cleaved mica cleaned by UV-ozone for 3 minutes. After 5 min of incubation, excess unbound amyloid was washed with 20 μL DI water 3-4 times. The sample was then dried under gentle $N_2$ gas until water evaporated and stored in a desiccator at room temperature prior to AFM imaging.

*Supported lipid bilayers (SLBs) for AFM.* Individual chloroform stocks of 1,2-dipalmitoyl-sn-glycero-3-phosphocholine (DPPC), 1-palmitoyl-2-oleoyl-sn-glycero-3-phosphocholine (POPC) and cholesterol (Sigma-Aldrich, Canada) were created (10 mg/mL each). Lipids were mixed by volume in a mass ratio of 4:4:2 and were then dried with $N_2$ gas to form a thin film. The lipid film was resuspended in NaCl solution and 100 mM Treh – NaCl solution, respectively, to reach a final concentration of 1mg/mL. The final lipid vesicle solution was prepared by alternating sonication and stirring for 15 minutes, respectively, until the solution became clear. To form an SLB, 150 - 250 μL of the vesicle solution was deposited on freshly cleaved mica that was previously treated with UV-ozone for 3 minutes and incubated for 30-40 minutes. After incubation, the SLB was washed with 8 – 15 mL DI water to remove excess solutions and imaged in DI water.

***SLBs incubated with Aβ1-42 for AFM.*** Once SLBs were formed as described, appropriate amounts of the 110 μM Aβ1-42 aliquot solution were added to SLB gently to a final concentration of 5 μM in the liquid cell. The SLB was incubated for 1 hour with Aβ1-42 prior to imaging.

***AFM imaging.*** NanoWizard II (JPK/Bruker, US) AFM instrument was used in intermittent contact mode to image Aβ1-42 oligomers in air and SLBs in liquid. PPP-NCH probes (Systems for Research (SFR), Canada) with spring constant 42 N/m were used to image dry Aβ-1-42 samples in air at resonance frequency of 293 kHz. QP-Bio-AC probes (SFR) with spring constant 0.3 N/m were used to image SLBs with and without Aβ1-42 in water at the resonance frequency of 25 – 35 kHz. Imaging for SLBs with Aβ1-42 began only after a pre-incubation period of 1 hour with Aβ1-42. Images were processed by JPK Data Processing software 5.1.4.

## RESULTS AND DISCUSSION

**Black Lipid Membrane Study of the Effect of Trehalose on Amyloid-Membrane Interactions**

In this study, we used BLM recordings to elucidate how membrane permeability is altered by the presence of amyloid and trehalose. The BLM electrophysiological technique records ion current fluctuations as the result of protein interactions with the lipid membrane suspended over the micrometer pore separating two compartments of a liquid chamber connected with two electrodes (Figure 2(a)). The suspended membrane acts as a capacitor with an extremely low conductance. A stable suspended membrane blocks the ion current efficiently with the initial current readout very close to 0 pA. When pore-forming proteins are added, they bind to the suspended membrane, interfere with membrane integrity, and induce ion leakage, as indicated by Figure 2(b) (82).

Using the BLM technique, we performed two types of measurements: first, we performed voltage dependence tests, which measure membrane conductance (current) versus voltage to evaluate the stability of membranes at various voltages applied with and without Aβ or trehalose. Currents were recorded in a series of stepwise voltages from -120 mV to 120 mV in increments of 10 mV. The slope of the curves plotted by current (I) vs voltage (V) in Figure 3 and Figure 4 indicated the membrane conductance. Secondly, we recorded ion currents at constant voltage over time to study the effect of amyloid binding and trehalose protection and evaluate the type of membrane damage observed. Single-event currents were recorded and analyzed at a constant voltage, ranging from 60 to 175 mV. This method was previously used to evaluate the single events and associating conductance caused by the interaction of Aβ1-42 with membranes (79).

## 1. *BLM Current/Voltage (I/V) Dependence Tests*

Current/Voltage (I/V) dependence tests are commonly used to evaluate membrane permeability and electric properties by measuring membrane conductance via stepwise voltages (83). Here, we used voltage-dependence measurements to evaluate the integral quality of membranes in buffers with and without trehalose, as well as to measure the membrane permeability before and after Aβ1-42 damage. This I/V relationship can also show the dependency of membrane conductance with respect to alterations in applied voltage.

*Effect of trehalose on lipid membrane conductance*

Considering the intrinsic ion leakage of lipid bilayers and ion currents through protein ion channels, the total currents across the membrane can be given by Eq.1 (84),

$$I = (g_L + g_p P_{open})(V - E_0) \qquad (1)$$

Where $g_L$ is membrane leakage conductance, $g_p$ is pore conductance, $P_{open}$ is probability that ion channels open, $V$ is applied voltage, $E_0$ is Nernst potential, which is equal to zero since the ion concentrations on both sides of membranes are identical (84).

Since there were no protein-induced events in the amyloid-free membranes and no ion concentration gradient, both $g_p P_{open}$ and $E_0$ were expected to be zero in Eq.1. Thus, the total current was attributed to the intrinsic leakage current and could be given by Eq.2 (84),

$$I = g_L V \qquad (2)$$

Figure 3 shows that the averaged I-V profile of untreated membranes fitted to this equation, demonstrating that the membrane conductance, $g_L$, was voltage-independent and constant.

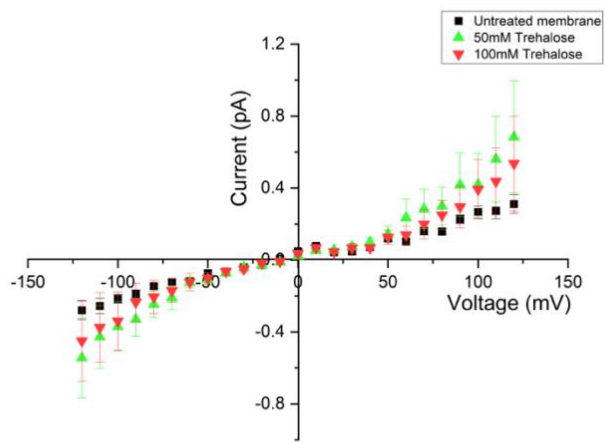

**Figure 3. Current/Voltage (I/V) dependence curves to estimate integral membrane permeability in buffers with and without trehalose by voltage dependence measurements.** Currents across each membrane were measured at -120 mV to 120 mV in increments of 10 mV. Black square: model lipid membranes tested in the 150 mM KCl – 20 mM HEPES salt buffer without trehalose; green triangle: membranes tested in 50 mM trehalose salt buffer; red triangle: membranes tested in 100 mM trehalose salt (150 mM KCl and 20 mM HEPES) buffer.

The I/V curves of membranes formed in pure salt buffer without trehalose (Figure 3, black square) followed the linear relationship of Ohm's Law with an average membrane conductance of 2.02 ± 0.08 pS (mean ± SEM., n = 10), which confirmed the inherent membrane stability and low ion permeability of our chosen lipid model. The symmetric and linear I-V relationship not only established that an even distribution of mixed lipid molecules was present on both sides of the cuvette aperture (membrane formation) but also demonstrated the constant and voltage-dependent nature of the resulting membrane conductance. Despite the theoretical notion that pure lipid membranes are impermeable to ions, the experimentally generated black lipid membranes showed a slight current due to the thermodynamic fluctuation of lipids—a process thought to trigger lipid pores (84). However, the low conductance of the untreated lipid bilayers (2.02 ± 0.08 pS) established that this fluctuation had minimal effects on overall membrane stability.

The I-V profiles of membranes formed in 50 mM and 100 mM trehalose, however, revealed the sugar's ability to alter the electric properties of amyloid-free membranes (Figure 3). These alterations were observed as small changes in conductance at lower applied voltages until the applied voltage reached |V| > 60mV, wherein the current exponentially increased with each

measured increase of applied voltage. The resulting I-V profile of trehalose membranes showed a deviation from the expected linear relationship of pure membranes when $|V| > 60mV$ were applied and instead resulted in an exponential relationship. Therefore, trehalose seemed to slightly reduce the robustness of treated membranes against higher voltages. In this case, the linear-fitted membrane conductance in both 50 mM trehalose (Figure 3, green triangle) and 100 mM trehalose (red triangle) was slightly higher than the trehalose-free membrane (black square). Specifically, the membranes incubated in 50 mM trehalose showed a larger magnitude of current increases and higher conductance ($2.48 \pm 0.14$ pS, n=12) as $|V| > 60mV$ compared to both untreated membranes ($2.02 \pm 0.08$ pS, n=10) and 100 mM trehalose treated membranes ($2.04 \pm 0.16$ pS, n = 10). This indicated that membranes in 50 mM trehalose buffer were the least robust to voltage changes, and therefore had the least stability, when compared to trehalose-free or 100mM trehalose buffers. The instability at low concentrations of trehalose can be explained by trehalose-lipid interactions.

Trehalose forms hydrogen bonds with the head groups of phospholipids and replaces water molecules at the lipid-solvent interface, thereby decreasing membrane dipole potential (66, 85, 86). As mentioned in a thermodynamic study by Villarreal *et al.*, both the trehalose-membrane electrostatic interaction and the substitution of water binding to lipid head groups by trehalose molecules are essential factors for the establishment of effective trehalose-lipid binding (66). As more trehalose binds to the membrane, water molecules near the lipids are replaced and some trehalose buries into the hydrophilic region of lipids (66). Given that trehalose is polar molecule, the accumulation of the sugar enhances the charge distribution not only at the lipid-solvent hydrophilic interface but also within the hydrophobic cores (66). Consequently, the displacement and dehydration process alter the thermodynamic environments and change electrostatic properties of lipid membranes.

Therefore, to ensure effective protection of these membranes, an adequate number of trehalose molecules and trehalose-lipid hydrogen bonding must occur to trigger the vital environmental and biophysical changes to these lipid membranes. In 50 mM trehalose, although membrane electric properties were altered, confirming the trehalose-lipid interaction, the number of trehalose molecules was insufficient to establish an effective barrier near the bilayer surface and instead reduced membrane stability. In 100 mM trehalose, the higher number of trehalose molecules was more effective at binding to lipid headgroups and forming a protective barrier, as indicated by the lower and more stable membrane conductance at high voltages. It is also possible that the massive accumulation of trehalose surrounding the lipids reduced the thickness of membranes. Molecular simulation studies show trehalose can expand the lipid molecular area and decrease the thickness of phospholipid bilayers in DPPC and POPC membranes (64, 87, 88). Moreover, the homogeneity of bilayer thickness is influenced by the trehalose/water ratio. Low concentrations of trehalose can trigger a heterogeneous distribution of bilayer thickness that might lead to the instability of bilayers, while higher concentrations can make the bilayer thickness distribution more uniform (87). This phenomenon may explain why moderate membrane instability appeared in 50 mM trehalose treated membranes instead of those treated with 100 mM trehalose. Overall, the moderate changes to membrane electrostatic properties had little effect on the stability of membranes suspended in trehalose buffer, according to the membrane conductance values obtained from the I/V profiles (Figure 3).

*Effect of Aβ on lipid membrane conductance*

Next, we added Aβ1-42 to suspended membranes and performed BLM recordings. The I-V curves generated by voltage-dependence measurements showed that the membranes incorporated with Aβ 1-42 had different electric characteristics when in the presence of trehalose compared to salt buffer alone (Figure 4). Currents across membranes exposed to Aβ1-42 without trehalose (green dot) increase linearly from – 60 mV to + 60 mV, the membrane conductance was 154.45 ± 14.11 pS; as |V| > 70 mV, current flow increased abruptly and lead to a larger membrane conductance 431.36

± 35.92 pS. The integral membrane conductance for these membranes (from -120 mV to 120 mV) is 168.84 ±17.02 pS (n = 6).

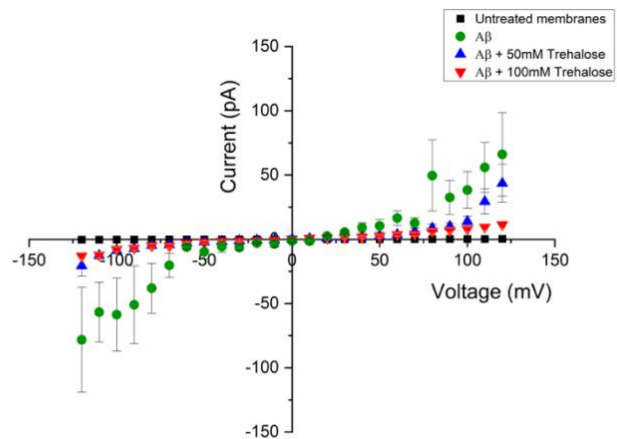

**Figure 4. I/V curves to estimate membrane conductance in the presence of Aβ(1-42)**. Membranes were suspended in three buffers (pure salt buffer, 50 mM trehalose salt buffer and 100 mM trehalose salt buffer) and incubated with 5μM Aβ1-42 prior to recording at -120 mV to 120 mV in increments of 10 mV. Black square: membranes in the salt buffer; Green dot: membrane in salt buffer incubated with Aβ1-42. Blue triangle: Aβ1-42 and 50 mM trehalose salt buffer. Red triangle: Aβ1-42 and 100 mM trehalose salt buffer.

In agreement with previous literature (82), our results demonstrated that Aβ aggregates in trehalose-free membranes induced a high degree of current fluctuation and dramatically altered membrane conductance specifically when voltage amplitudes exceeded 70 mV ($|V| > 70$ mV). This increase of current magnitude quickly became disproportional to the controlled increase of the applied voltage (Figure 4, green dot). Towards the limits of the voltage ranges ($|V| > 70$ mV) we observed membrane damage induced by Aβ1-42 oligomers as indicated by this stronger response to voltage changes and greater current flux. Compared to the untreated membrane, the membrane conductance was dramatically increased from 2.02 ± 0.08 pS (n = 10, black square) to 168.84 ±17.02 pS (n =6, green dot), as a consequence of Aβ1-42 damage.

I/V profiles of Aβ1-42-exposed membranes confirmed that Aβ1-42 oligomers interacted with lipid membranes, causing increases in membrane permeability to ions and ultimately leading to the disruption of membrane integrity (Figure 4). Notably, when 80 mV was applied to membranes with amyloid present, there was an abrupt current increase (green dot); this increase in applied voltage may promote ion channel formation or larger pore sizes. Since Aβ is negatively charged at physiological pH (pH 7.4) (89), the application of higher voltages could increase the mobility of ions as well as charged Aβ, which in turn promotes Aβ insertions in the membrane. A similar scenario was observed in the cellular membranes of SH-SY5Y and cortical neuron cells, where a large current fluctuation caused by Aβ occurred at – 80 mV, which is close to the cells' resting potential (24). This study could not exclude the possibility that Aβ causes robust membrane instability during insertion events when higher voltages are applied. Both Aβ damage and membrane deformation by electrostriction under high voltages can lead to dramatic conductance increases and promote membrane disruption.

*Effect of trehalose on amyloid-induced membrane conductance*

In order to evaluate the protective effect of trehalose, we added trehalose solution in combination with amyloid to suspended membranes. I/V profiles of Aβ1-42-exposed membranes with trehalose are shown in Figure 4 (red and blue triangles). Although amyloid-induced damage was still observed, it was greatly reduced by trehalose. Specifically, the Aβ1-42-induced currents were reduced in the presence of 50 mM trehalose at $|V| < 70mV$, where membrane conductance was $43.09 \pm 2.23$ pS. However, it still displayed exponential growth at $|V| > 70mV$ (blue triangle, n = 7) with a membrane conductance of $99.89 \pm 11.67$ pS, generating the integral membrane conductance of $46.71 \pm 3.40$ pS. Conversely, the membranes in 100 mM trehalose buffer demonstrated very little fluctuation in the Aβ-induced currents at $|V| > 70mV$ (red triangle, n = 6). In 100 mM trehalose, membrane conductance was $51.55 \pm 0.00$ pS at $|V| < 70mV$ and was $84.10 \pm 6.00$ pS at $|V| > 70mV$; the integral membrane conductance from -120 mV to 120 mV was $66.36 \pm 3.44$ pS. Despite the persistence of Aβ-membrane interactions in the presence of trehalose, both concentrations (50mM, 100mM) of trehalose decreased the overall membrane conductance and the amplitude of Aβ(1-42)-induced current flow as compared to the trehalose-free buffer. The shape and trends of the I-V profiles observed in all treatments demonstrated exponential growth at higher voltage (Figures 3 & 4). However, those membranes treated with trehalose in the presence of Aβ showed a lower membrane conductance, suggesting that trehalose suppressed current flow across Aβ-damaged membranes, and thus acted as a protective agent against amyloid toxicity.

While Aβ1-42 ion channels/insertion events were still seen in the presence of trehalose, the recorded current flux and membrane conductance in 50 mM and 100 mM decreased remarkably from 168.84 pS to 46.71 pS and 66.36 pS, respectively (Figure 4). The lowered membrane conductance can be explained by two potential mechanisms: 1) Trehalose suppressed the conformational change of Aβ1-42 as it inserts into suspended membranes by limiting pore diameters therefore limiting ion flux created by Aβ oligomers. Previous studies show that Aβ undergoes conformational changes during insertion into lipid bilayers (90). The binding of trehalose to lipid membrane headgroups could potentially restrict this membrane-induced conformation change during insertion and consequently reduce the original ion-channel pore sizes formed by Aβ. 2) The lipid dehydration by trehalose and the generation of this protective barrier maintained membrane stability when exposed to Aβ1-42 by avoiding membrane collapse during peptide insertion events. In addition to generating ionic pores, Aβ1-42 can also damage membranes by lowering bilayer dielectric barriers, consequently increasing membrane conductance (91). Once bound, trehalose molecules worked as a protective barrier and prevented Aβ access to the membrane, hindering the considerable drop of dielectric barriers and stabilizing membrane structure. Consequently, trehalose-containing lipid bilayers became less permeable to ions compared to sugar-free membranes when in the presence of Aβ. Although trehalose failed to block Aβ1-42-membrane interactions completely, membrane conductance during these interactions was significantly reduced at higher trehalose concentrations. Specifically, the reduction of membrane conductance in the 100 mM trehalose buffer showed the alleviation of amyloid oligomer-induced damage to lipid membranes as a result of the sugar's presence.

## 2. Single-event Recordings of Induced Membrane Damage (Current Versus Time)

The single-event recordings were conducted to estimate the instantaneous damage level to membranes after amyloid exposure. A constant voltage in the range of 60 to 175 mV was applied, and the current fluctuation was recorded over time while the suspended membranes were exposed to an amyloid solution, with and without trehalose present. Current vs time plots are shown in Figure 5 and display several types of current fluctuations: single spike currents and broader collective current bursts. Figure 5(a) shows representative current flux caused by Aβ1-42 insertion into a suspended membrane at 120 mV.

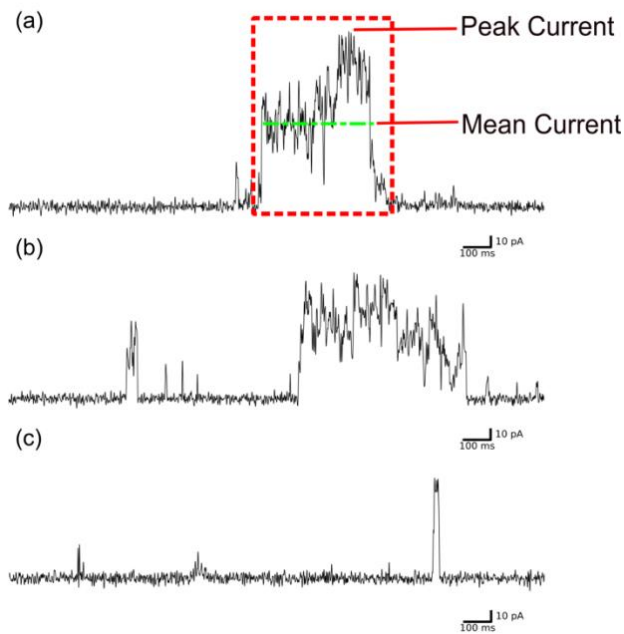

**Figure 5. Amyloid-membrane interactions and current fluctuation induced by Aβ damage to membranes respect to the time.** Representative current flux insertion events caused by Aβ1-42 in (a) 0 mM trehalose, (b) 50 mM trehalose, (c) 100 mM trehalose buffers at 120 mV. The baseline was 0 pA. The peak current (pointed by a red line) was taken at the maximum value of current in a single event as the current increased. The mean current (green dash) was obtained by averaging the current values within a single event, from where the current dramatically increased to where it dropped to the baseline.

Figure 5(b) shows the currents observed due to Aβ1-42 insertion into the suspended membrane in the presence of trehalose at 50 mM and Figure 5(c) at 100 mM trehalose. The maximum (peak) and mean conductance correlating to peak and mean currents of single events in Figure 5 were analyzed; the results of the analysis are shown in Figure 6 and Table 1.

A combination of the single-spike event and broad distribution of ion currents in Figure 5 is in agreement with the Aβ-membrane hypothesis which states that amyloid is capable of forming various types of defects in the membrane, such as single well-defined ion channels, as well as larger pores of various sizes and lipid dissociation due to detergent-like effects (23, 92), which in turn may be attributed to the variety of sizes and conformations of Aβ aggregates, interacting with membranes (23). As shown in Figure 1(a), in addition to forming ion channels, Aβ aggregates may penetrate the membrane directly or adhere to the membrane surface (23).

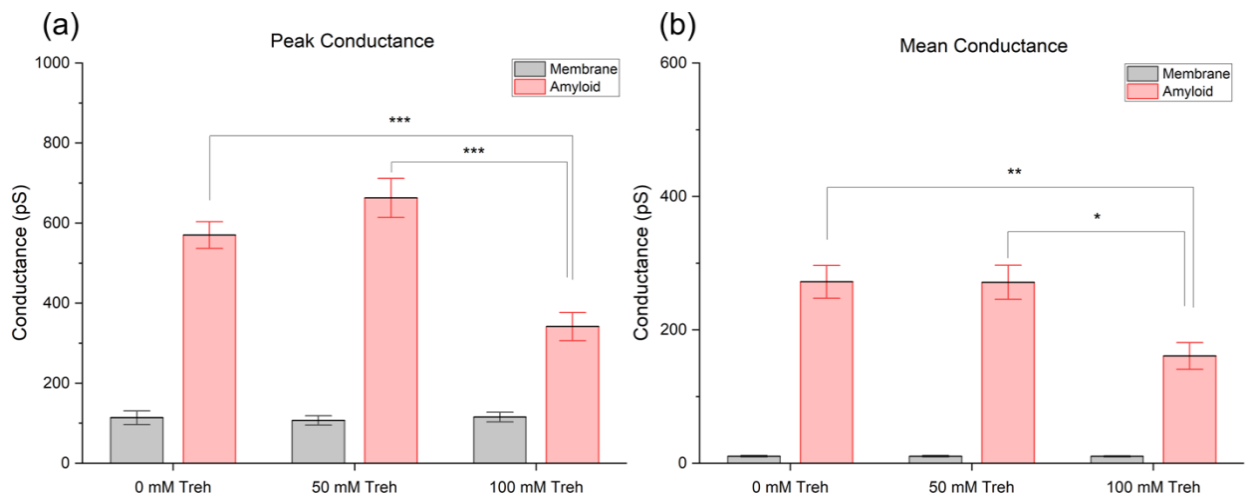

**Figure 6. Ion current dependence vs time of incubation with amyloid, and its dependence on trehalose (Treh) concentration at constant voltage. (a) Peak and (b) Mean conductance of single events allows evaluation of instantaneous amyloid destruction to membranes.** Currents of single amyloid insertion events in membranes exposed to 5 μM Aβ1-42 were recorded and compared to those suspended in Aβ-free salt buffers. The peak conductance was calculated from dividing the maximum value of a single insertion event (AB insertion induced current increase) by the corresponding applied voltage. The mean conductance was calculated by the average value of single current increase caused by amyloid-membrane interaction divided by corresponding voltages. Black: amyloid-free membranes; Pink: Aβ1-42-interacted membranes. Left column: salt buffer without trehalose; middle: 50 mM trehalose salt buffer; right: 100 mM trehalose salt buffer. The values of conductance (peak and mean) for each condition are in Table 1.

**Table 1. Peak and Mean conductance values corresponding to each condition in Figure 6.**

| Condition | Sample size (N) | Peak Conductance (pS) | Mean Conductance (pS) |
| --- | --- | --- | --- |
| **Membrane** | 48 | 113.73 ± 17.12 | 10.55 ± 1.15 |
| **Membrane +Aβ1-42** | 213 | 570.11 ± 33.39 | 272.06 ± 24.58 |
| **Membrane + 50 mM Treh** | 53 | 107.07 ± 11.80 | 10.55 ± 1.14 |
| **Membrane + 50 mM Treh + Aβ1-42** | 118 | 662.85 ± 48.78 | 271.35 ± 25.56 |
| **Membrane + 100 mM Treh** | 56 | 115.56 ± 12.06 | 10. 35 ± 0.88 |
| **Membrane + 100 mM Treh + Aβ1-42** | 114 | 365.67 ± 33.98 | 160.21 ± 20.18 |

During the electrophysiological recordings, single events of current escalation observed in the real-time readout represented current leakage and decreased membrane stability due to short-term interactions between Aβ1-42 oligomers and the suspended membranes. The duration of current fluctuation has various lifetimes (Figure 2 & 5)—from a few milliseconds to thousands of milliseconds—indicative of both varying methods of Aβ1-42 interaction with membranes, as well as the diverse size of Aβ1-42 aggregates. The level of instantaneous conductance increase caused by the proteins correlates to the size of pores in the membrane (82). Due to this, peak conductance and mean conductance could be derived from a single signal, and therefore, both parameters were considered for the estimation of membrane damage caused by Aβ1-42. Peak conductance uses the single highest measurement of current recording to evaluate maximum membrane damage after Aβ1-42 insertion while mean conductance was used to evaluate the average destruction level as the protein-induced damage occurred. The peak and mean conductance of membranes in amyloid-free buffers with and without trehalose demonstrated a low level of current flow indicating membrane stability under these conditions (Figure 6). However, the addition of Aβ1-42 to the membrane and subsequent interaction triggered current leakage, indicated by the observed increase in instantaneous currents and conductance in all three buffer types (Figure 5). The 50 mM trehalose demonstrated the levels of current leakage across the membrane similar to pure salt buffers without trehalose (Figure 6), comparing the peak and mean conductance (one-way ANOVA; $p > 0.05$); suggesting that the 50 mM concentration has failed to protect the membrane.

Conversely, the higher concentration of trehalose (100 mM) significantly enhanced the resistance of membranes to amyloid toxicity and maintained stability, indicating membrane protection from peptide damage. The peak conductance of membranes incubated with amyloid (570.11 ± 33.39 pS, N = 213) decreased significantly in the 100 mM trehalose buffer (365.67 ± 33.98 pS, N = 114), according to the results of a one-way ANOVA ($p < 0.01$, DF = 444) and Tukey's test ($p < 0.001$), as shown in Figure 6(a). Moreover, the mean conductance of current fluctuation in single events was also lowered in 100 mM trehalose (Figure 6(b), one-way ANOVA: $p < 0.01$, DF = 444; Tukey's test, $p < 0.05$). This indicates that 100 mM trehalose is capable of suppressing Aβ1-42-induced current flux across the membrane. It is also worth mentioning that the Aβ-membrane insertion events were observed with a lower event occurrence in both trehalose solutions according to the number of effective events (sample size in Table 1). While 50 mM trehalose failed to effectively inhibit amyloid damage, high concentrations of trehalose (100mM) not only reduced the insertion of Aβ proteins into membranes but also reduced damage, thereby preserving membrane stability. The difference in results between sugar concentrations (50mM vs 100mM) indicated that the protective nature of trehalose may be concentration-dependent.

Previously published molecular dynamics simulations by Izmitli *et al.* showed that trehalose shortened the retaining time of Aβ peptide in lipids and encouraged the rapid insertion (67, 93). Trehalose can reduce the contact period for lipid membranes and Aβ thereby reducing the lateral damages from amyloid proteins. The fast movement of Aβ across the membrane may also decrease the loss of lipid molecules due to protein retention and provide the membrane more time to recover the integrity from Aβ in the presence of trehalose. In addition, the simulation studies elucidated that the trehalose subphase quantitively decreased the insertion of Aβ to lipids, compared to the water subphase (67), which is consistent with our observation in experiments -- the occurrence of single insertion events was lowered in trehalose solutions (Table 1). While we could not exclude the impact of trehalose on protein mobility, we deduced that trehalose at high concentrations maintained the stability of membranes after amyloid destruction and reduced the probability of membrane–protein interaction by preventing Aβ1-42 access to the membrane. Thus, high concentrations of trehalose (100mM) can create a sufficiently protective barrier and reduce ion flux for suspended membranes against the interference and damage induced by Aβ1-42, thus supporting our hypothesis illustrated in Figure 1.

**Atomic Force Microscopy.**

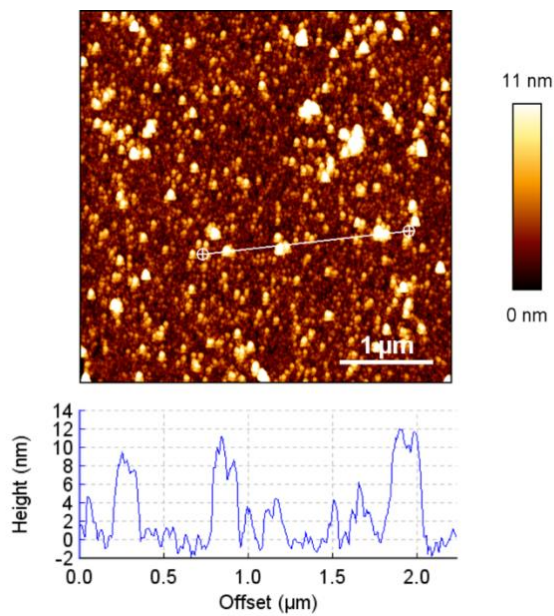

**Figure 7. Aβ oligomers on mica imaged by AFM.**

Next, we used AFM imaging to visualize and confirm formation of Aβ1-42 oligomers (Figure 7) at the same conditions as they interacted with suspended membranes in BLM experiments, and to visualize the membrane damage in lipid membranes as a result of amyloid binding to the membranes (Figure 8).

To confirm the aggregation state of the peptide used in BLM experiments, the aliquot of Aβ1-42 was allowed to incubate on mica for 5 minutes, rinsed gently, dried with nitrogen and imaged with AFM in air.  As can be seen in Figure 7, the amyloid peptide formed small spheric oligomers, with most ranging between 2-10 nm in height according to AFM cross-sections, with a few larger particles exceeding 10 nm present. The sizes of Aβ1-42 oligomers were aligned with the range of Aβ oligomers reported by previous studies (94).

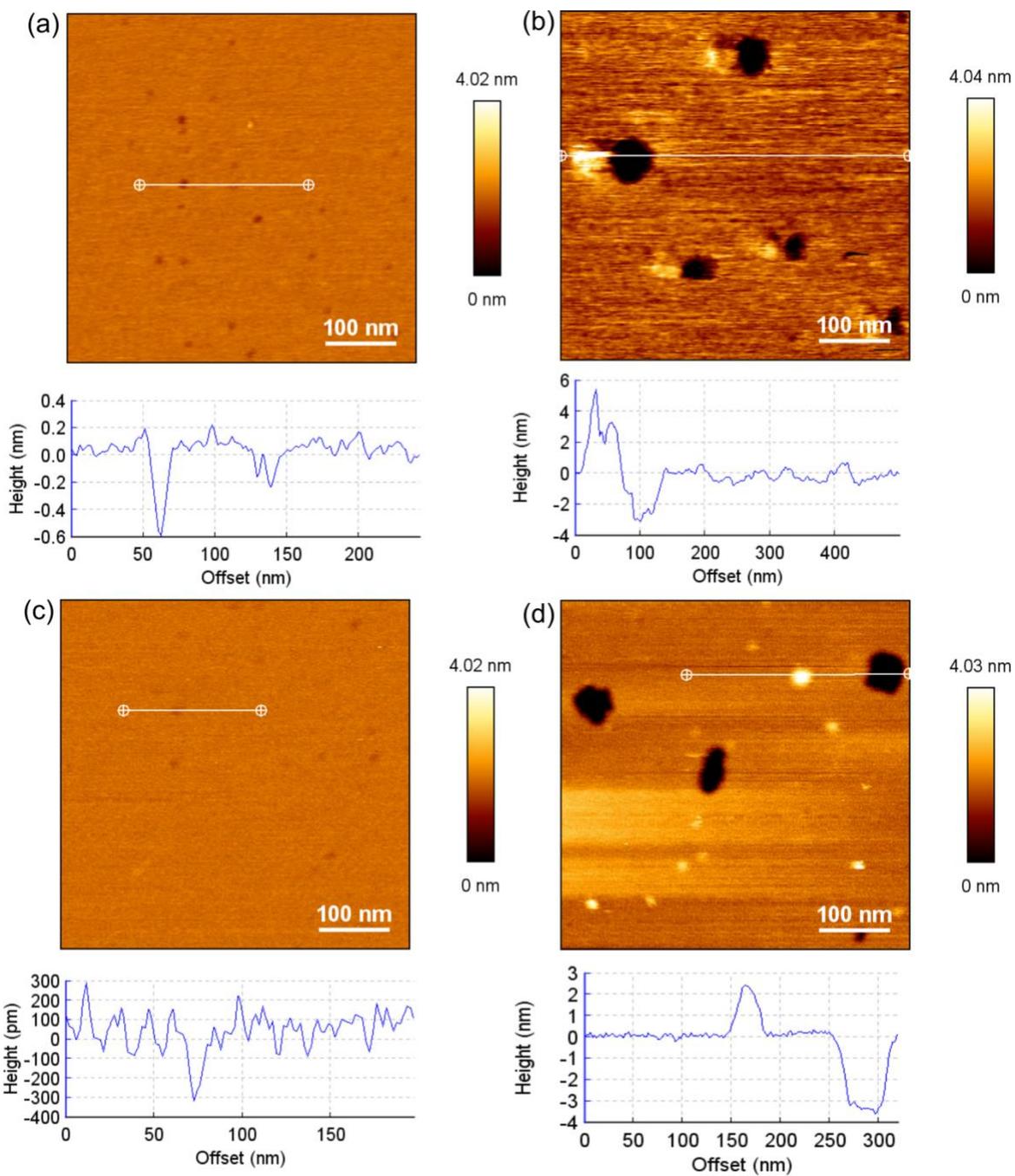

**Figure 8. Representative AFM images of lipid membranes and corresponding cross-section profiles.** (a) supported lipid bilayers (SLB) composed of DPPC, POPC and cholesterol in 4:4:2 weight ratio. (b) SLBs incubated with Aβ1-42 at 5 μM for 1h, displaying the Aβ-induced pore in SLBs. (c) Intact SLB made in 100 mM trehalose (Treh-SLB), imaged in water. (d) Trehalose-SLB incubated with Aβ1-42 oligomers.

Figure 8(a) shows an AFM topographical image of the model membrane surface with its associated cross-section (4:4:2, DPPC: POPC: Cholesterol by weight). The image and cross-section indicate that the membrane surface was relatively flat with smaller, dispersed lower (dark) domains with a height difference of ~ 0.6 nm. Similar scenarios appeared in the cholesterol-enriched membrane in previous studies (95, 96). The abundance of cholesterol in the membrane could be the reason behind the absence of explicit phase separation and the formation of nanodomains. The cholesterol tends to condense with phospholipids, which can make the fluidic POPC more rigid, while it reversely fluidizes the gel-phase DPPC at room temperature (97). This can attenuate the phase separation due to the difference of lipid components in the SLB. After imaging the membrane, Aβ1-42 was added to the sample at a final concentration of approximately 5 μM. After incubation with 5 μM Aβ1-42 oligomers for 1 hour, Aβ-induced pores of varying lateral sizes were observed on the membrane surface (Figure 8(b)). The higher features with ambiguous edges next to each pore indicate Aβ1-42 embedding into the bilayer. This phenomenon is consistent with observations in previous AFM studies that show pore formation in model membranes as a result of small-size Aβ oligomers (98). In Figure 8(b), the height of the Aβ aggregates embedded in the membrane was approximately 4.5 nm. These aggregates induced nearby pore formation in the lipid membrane, indicated in the cross-section as a depression with a height of 3 nm. Aside from these pore-forming Aβ aggregates, some small Aβ penetrated the lipid bilayers without causing pores (Figure S1 in the Supplementary material). The AFM images confirmed the variety of Aβ-membrane interaction, in aligned with BLM recordings (Figure 5).

AFM images of the Trehalose-SLB also showed a flat membrane with small nanodomains, elucidating a subtle change in membrane morphology (Figure 8(c)) compared to SLB without trehalose (Figure 8(a)). This observation is consistent with previous neutron diffraction studies by Kent *et al.*, which confirmed that trehalose does not change the structure of lipid membranes in the hydration environment (99). Additionally, recent simulation studies also reported that trehalose was more likely to maintain the original phase separation and fluidity of lipid membranes instead of changing them (100). Therefore, it is rational that the original morphology of the lipid membrane was preserved in trehalose. Trehalose-SLB were incubated with Aβ1-42 oligomers for 1h and the corresponding topographic images were obtained, as shown in Figure (8(d)). Like SLB with Aβ1-42 in Figure 8(b), the Aβ-treated Trehalose-SLB also featured pore formation with Aβ1-42 proteins embedded in the surface. In conclusion, AFM results indicated that the Aβ1-42 oligomers can induce pore formation in SLBs without trehalose as well as in Trehalose-SLBs, consistent with the BLM results where the Aβ-induced ion leakage could be detected in both suspended membranes with and without trehalose.

*Membrane Protection Mechanism of Trehalose*

In our study, we hypothesized that trehalose protection mechanism works on a membrane level and helps maintain integrity of lipid membranes during interaction with Aβ1-42. AFM images of SLBs showed that the Aβ1-42 damage to membranes existed and membrane disruption occurred in all conditions (with and without trehalose), in alignment with the observation from the single event recordings by BLM. Whereas, according to the voltage dependence measurements, the lowered membrane conductance in 100mM trehalose indicated that trehalose maintained membrane integrity even after damage by Aβ1-42. Additionally, the reduction of instantaneous conductance in trehalose solutions showed that trehalose stabilized the membrane as the insertion events occurred. Hence, our AFM and BLM studies reveal that trehalose can significantly protect membranes. It works on a membrane level instead of blocking Aβ1-42 in solution.

Currently, there is a lack of experimental evidence regarding the interaction between trehalose, Aβ oligomers, and membranes. Previous molecular dynamics simulation studies only report the interactions of monomeric Aβ with membranes (93), and while these studies demonstrate the ability of trehalose to mitigate the conformational change of Aβ during membrane insertion (67), they focus on a single Aβ peptide rather than toxic Aβ oligomers.

Another possible explanation suggests that trehalose changes the packing of lipid molecules (100) and strengthens the bilayer resistance to amyloid oligomer damaging effects, thereby attenuating Aβ-induced membrane damage and maintaining membrane stability. Further research is necessary to fully understand the impact of trehalose on Aβ oligomer insertion into lipid membranes, both theoretically and experimentally.

The current therapeutic protection of trehalose in neurodegenerative disorders, including AD, is mainly attributed to the activation of autophagy, evidenced in animal models (63, 101-103). Nevertheless, while autophagy induced by trehalose can occur in various regions in biological systems (63), the role of trehalose in nervous systems is still unclear. Although trehalose is already found in blood plasma, kidney and liver, it is still unknown whether trehalose can be transported across the blood-brain barrier (BBB) into human brains. Martano et al. found endogenous trehalose, and correlated trehalase in the hippocampus and cortex of rodent brains (104). The trehalose can be produced by primary astrocytes and contributes to neuronal arborization and morphology development (104), implying nervous function of trehalose in vertebrates' brains. Our studies demonstrate a potential mechanism of neuroprotection conducted by trehalose in AD – trehalose can attenuate the Aβ-mediated membrane disruption, thereby protecting neuronal membranes. Given the complexity and diversity of lipid composition within brains, it is worth further investigation on the ability of trehalose to protect more complex membrane systems.

Currently, clinical trials of trehalose in AD patients are undergoing and mainly administrated by oral intake and infusion (63). The improvement of the therapeutic usage of trehalose is being realized by chemical modification of trehalose, which can reduce the dosage, enhance the trehalose penetration into neurons, elevate therapeutic effect and mitigate trehalose-induced adverse effects, etc. (63). In addition, the incorporation of trehalose in drug delivery systems (e.g. liposome, nanoparticles) may assist transportation of trehalose across BBB and increase its specificity (105). While these relevant studies shed light on trehalose as a drug candidate for neuroprotection in AD, understanding the molecular mechanisms of trehalose protection in the model membranes can benefit the exploration of interdisciplinary pharmaceutical strategies for neurodegenerative diseases.

**CONCLUSION**

In summary, our BLM and AFM data demonstrate that trehalose can protect model lipid membranes from the damaging effects of Aβ1-42 oligomers. The results of voltage dependence tests and single-event measurements performed in the BLM system indicate that trehalose changed the electric and biophysical properties of pure membranes and amyloid-exposed membranes. The lowered membrane conductance in the trehalose-Aβ environment verified the ability of trehalose to protect model neuronal membranes against Aβ1-42 toxicity. This study provides new insights into understanding the molecular mechanism of protective agents against Aβ neurotoxicity with a focus on the properties of the membrane itself and thus may promote the development of novel directions for preventive strategies in Alzheimer's Disease. Thus, trehalose membrane protection against amyloid toxicity is worthy of more exploration in cellular studies and animal models in future.

## AUTHOR CONTRIBUTIONS

**YX:** designed research, performed experiments, analyzed data, and wrote the manuscript. **CTF:** contributed to designing research and writing manuscript**, ZL:** designed and supervised the research, edited and finalized manuscript.

## DECLARATION OF INTERESTS

The authors declared that there is no conflict of interests.

## ACKNOWLEDGEMENTS

The authors acknowledge funding and support from Natural Sciences and Engineering Research Council of Canada (NSERC), Canada Foundation for Innovation (CFI), Ontario Research Fund (ORF) and the Waterloo Institute of Nanotechnology Fellowship to CTF.

## SUPPORTING CITATIONS

Reference (106) appears in the Supporting Material.

# Protective Effect of Trehalose Sugar on Amyloid-Membrane Interactions using BLM Electrophysiology

Yue Xu[1], Carina Teresa Filice[2,3], Zoya Leonenko[1-3, *]

[1]Department of Physics & Astronomy, University of Waterloo, Waterloo, ON, N2L 3G1, Canada;

[2]Department of Biology, University of Waterloo, Waterloo, ON, N2L 3G1, Canada;

[3]Waterloo Institute for Nanotechnology, Waterloo, ON, N2L 3G1, Canada

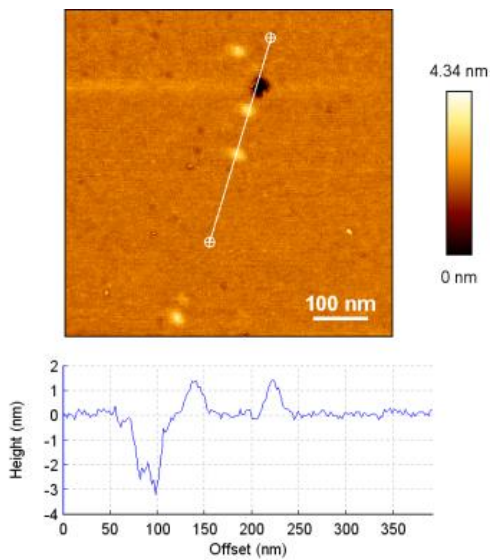

**Figure S1. Supported lipid bilayer (SLB) embedding with Aβ oligomers.**

Figure S1 shows SLB embedding with Aβ oligomers. The retention of Aβ peptides in lipid membrane was also observed in the study by Siniscalco *et al*. (1). The variety in membrane interaction could be due to the diverse sizes of Aβ1-42 oligomers as well as the retention time of Aβ within the membrane. Mrdenovic *et al.* perceived the process of pore expansion and enclosure as Aβ oligomers penetrated and fused membranes from 0 min to 91 min by time-lapse AFM (2). These AFM images (Figure S1, Figure 8(b)&(d) in the main text) confirm the Aβ-membrane interaction and explain the variety of Aβ-induced current flow observed in our BLM recordings.

## SUPPORTING REFERENCES